\documentclass[preprint]{elsarticle}
\usepackage{amsfonts}
\usepackage{amssymb,graphics,epsfig,subfigure}

\journal{Physics Letters B}

\begin{document}
\begin{frontmatter}
\title{Effects of dark energy on $P$-$V$ criticality of charged AdS black holes}
\author{Gu-Qiang Li}
 \address{Institute of Theoretical Physics, Zhanjiang Normal University, Zhanjiang, 524048, Guangdong, China}

\begin{abstract}
   In this Letter, we investigate the effects of dark energy on $P-V$ criticality of charged AdS black holes by considering the case of the RN-AdS black holes surrounded by quintessence. By treating the cosmological constant as thermodynamic pressure, we study its thermodynamics in the extended phase space. It is shown that quintessence dark energy does not affect the existence of small/large black hole phase transition. For the case $\omega_q=-2/3$ we derive analytic expressions of critical physical quantities, while for cases $\omega_q\neq-2/3$ we appeal to numerical method for help. It is shown that quintessence dark energy affects the critical physical quantities near the critical point. Critical exponents are also calculated. They are exactly the same as those obtained before for arbitrary other AdS black holes, which implies that quintessence dark energy does not change the critical exponents.

\end{abstract}
\end{frontmatter}

\section{Introduction}
     Charged AdS black holes share striking similarity with liquid-gas systems. Chamblin et al.~\cite{Chamblin1,Chamblin2} discovered that there exist first order phase transitions among Reissner-Nordstr\"{o}m-AdS (RN-AdS) black holes, whose critical behavior is analogous to the Van der Waals liquid-gas phase transition.

     Treating the cosmological constant as thermodynamic pressure, black hole thermodynamics in the extended phase space has gained more and more attention recently~\cite{Kastor}-~\cite{Decheng2} and the analogy between charged AdS black holes and liquid-gas systems has been further enhanced. Kubiz\v{n}\'{a}k et al.~\cite{Kubiznak} reconsidered the critical behavior of charged AdS black holes in the extended phase space. It was shown that there exists a critical temperature below which a first order small-large black hole phase transition occurs. In $P$-$T$ diagram there also exists a coexistence line terminating at the critical point. Critical exponents were calculated and shown to coincide with those of the liquid-gas systems. Moreover, it was stressed that the analogy is exact correspondence of physical quantities rather than a mathematical analogy in former literatures. Their work was soon elaborated~\cite{Gunasekaran}-\cite{Decheng2} by researchers all over the world. For a nice review, see Ref.~\cite{Altamirano3}.

     It is widely believed that our universe is expanding with acceleration. This opinion is supported by a variety of modern observational results \cite{Bachall}-~\cite{Sahni} and may be explained by the existence of dark energy, which makes up about 70 percent of our universe. Among the various dark
     energy model, quintessence is proposed as canonical scalar field with state parameter satisfying $-1<\omega_q<-1/3$ to explain the late-time cosmic
acceleration \cite{Fujii}- \cite{Ratra}. For a recent review, see Ref. \cite{Tsujikawa}. Black holes surrounded by quintessence have received considerable attention and their thermodynamics has been intensively investigated~\cite{Kiselev}-~\cite{Kuriakose3}. It would be of interest to probe whether dark energy influences the critical behavior of black holes in the extended phase space which we discussed in the former paragraph. To achieve this goal, we would like to investigate $P$-$V$ criticality of the RN-AdS black hole surrounded by quintessence in this Letter. To our best knowledge, thermodynamics in the extended phase space of the RN-AdS black hole surrounded by quintessence has been absent in literature yet. In this letter, we would first study its thermodynamics in the extended phase space and then probe how dark energy influences its critical behavior. We would also calculate its critical exponents. With the existence of quintessence, the equation of state in the extended phase space becomes more complicated and it would be a non-trivial task to solve the equations to get the analytic expressions of critical point. So we would solve a special case analytically and leave other cases to be solved numerically.

    The organization of this Letter is as follows. In Sec.\ref{Sec2}, thermodynamics of the RN-AdS black hole surrounded by quintessence will be studied in the extended phase space. In Sec.\ref{Sec3}, $P$-$V$ criticality of the RN-AdS black hole surrounded by quintessence will be investigated and the effect of quintessence dark energy will be discussed. In Sec.\ref {Sec4}, relevant critical exponents will be calculated. Conclusions will be drawn in Sec.\ref{Sec5}.

\section{Thermodynamics of the RN-AdS black hole surrounded by quintessence}
\label{Sec2}
The metric of the RN-AdS black hole surrounded by quintessence can be written as~\cite{Kiselev}
\begin{equation}
ds^2=f(r)dt^2-f(r)^{-1}dr^2-r^2(d\theta^2+\sin^2\theta d\varphi^2),\label{1}\\
\end{equation}
where
\begin{equation}
f(r)=1-\frac{2M}{r}+\frac{Q^2}{r^2}-\frac{a}{r^{3\omega_q+1}}-\frac{\Lambda r^2}{3},\label{2}\\
\end{equation}
where $\omega_q$ is the state parameter while $a$ is the normalization factor related to the density of quintessence. It is well known that for quintessence dark energy $-1<\omega_q<-1/3$, while for phantom dark energy $\omega_q<-1$. With the normalization factor $a$, the density of quintessence
can be expressed as
\begin{equation}
\rho_q=-\frac{a}{2}\frac{3\omega_q}{r^{3(\omega_q+1)}}.\label{3}\\
\end{equation}
Solving the equation $f(r_+) =0$, one can obtain the event horizon radius $r_+$, with which the mass of the black hole can be expressed as
\begin{equation}
M=\frac{r_+}{2}\left(1+\frac{Q^2}{r_+^2}-\frac{a}{r_+^{3\omega_q+1}}-\frac{\Lambda r_+^2}{3}\right).\label{4}\\
\end{equation}

The Hawking temperature can be derived as
\begin{equation}
T=\frac{f'(r_+)}{4\pi}=\frac{1}{4\pi}\left(\frac{1}{r_+}-\frac{Q^2}{r_+^3}+\frac{3a\omega_q}{ r_+^{2+3\omega_q}}-r_+\Lambda\right).\label{5}
\end{equation}

The entropy can be calculated as
\begin{equation}
S=\int^{r_+}_0\frac{1}{T}\left(\frac{\partial M}{\partial r_+}\right)dr_+=\pi r_+^2.\label{6}
\end{equation}

In the extended phase space, one can treat the cosmological constant as thermodynamic pressure and its conjugate quantity as
thermodynamic volume. The definitions are as follows
\begin{eqnarray}
P&=&-\frac{\Lambda}{8\pi},\label{7}
\\
V&=&\left(\frac{\partial M}{\partial P}\right)_{S,Q}.\label{8}
\end{eqnarray}
With Eqs. (\ref{4}) and (\ref{7}), the mass can be reexpressed as
\begin{equation}
M=\frac{r_+}{2}\left(1+\frac{Q^2}{r_+^2}-\frac{a}{r_+^{3\omega_q+1}}+\frac{8\pi P r_+^2}{3}\right).\label{9}
\end{equation}
Utilizing Eqs. (\ref{8}) and (\ref{9}), one can obtain the thermodynamic volume as
\begin{equation}
V=\frac{4\pi r_+^3}{3}.\label{10}
\end{equation}
Comparing Eqs. (\ref{6}) and (\ref{10}) with those of RN-AdS black holes~\cite{Kubiznak}, it is interesting to note that the existence of quintessence dark energy does not affect the expressions of both the entropy and the volume of RN-AdS black holes.

With the newly defined thermodynamic quantities, the first law of black hole thermodynamics in the extended phase space can be written as
\begin{equation}
dM=TdS+\Phi dQ+VdP+\mathcal {A}da,\label{11}
\end{equation}%
where $\mathcal {A}$ is the physical quantity conjugate to the parameter $a$. It is introduced to make the first law consistent with the Smarr relation and its physical meaning needs further investigation.

Utilizing Eqs. (\ref{9}) and (\ref{11}), one can obtain
\begin{eqnarray}
\Phi&=&\left(\frac{\partial M}{\partial Q}\right)_{S,P,a}=\frac{Q}{r_+},\label{98}
\\
\mathcal {A}&=&\left(\frac{\partial M}{\partial a}\right)_{S,Q,P}=-\frac{1}{2r_+^{3\omega_q}}.\label{99}
\end{eqnarray}

And the Smarr relation reads
\begin{equation}
M=2TS+\Phi Q-2VP+(1+3\omega_q)\mathcal {A}a.\label{100}
\end{equation}%
Note that the Smarr relation can also be derived from the first law by scaling argument as Ref.~\cite{Kastor} did.

\section{$P$-$V$ criticality of the RN-AdS black hole surrounded by quintessence}
\label{Sec3}
With Eqs. (\ref{5}) and (\ref{7}), the Hawking temperature can be reexpressed as
\begin{equation}
T=\frac{1}{4\pi}\left(\frac{1}{r_+}-\frac{Q^2}{r_+^3}+\frac{3a\omega_q}{ r_+^{2+3\omega_q}}+8\pi P r_+\right).\label{12}
\end{equation}
From Eq. (\ref{12}), one can easily derive the equation of state of the black hole as
\begin{equation}
P=\frac{T}{2r_+}-\frac{1}{8\pi r_+^2}+\frac{Q^2}{8\pi r_+^4}-\frac{3a\omega_q}{8\pi r_+^{3(1+\omega_q)}}.\label{13}
\end{equation}
Denoting $2r_+$ by $v$, one can obtain
\begin{equation}
P=\frac{T}{v}-\frac{1}{2\pi v^2}+\frac{2Q^2}{\pi v^4}-\frac{8^{\omega_q}\times 3a\omega_q}{\pi v^{3(1+\omega_q)}}.\label{14}
\end{equation}

The critical point can be derived through the following conditions
\begin{eqnarray}
\left.\frac{\partial P}{\partial v}\right|_{T=T_c}&=&0,\label{15}\\
\left.\frac{\partial ^2P}{\partial v^2}\right|_{T=T_c}&=&0.\label{16}
\end{eqnarray}
Utilizing  Eqs. (\ref{14}) and (\ref{15}), one can get
\begin{equation}
T_c=\frac{1}{\pi v_c}-\frac{8Q^2}{\pi v_c^3}+\frac{8^{\omega_q}\times 9a\omega_q(1+\omega_q)}{\pi v_c^{2+3\omega_q}},\label{17}
\end{equation}
where $T_c$, $v_c$ denote the critical Hawking temperature and critical specific volume respectively.
Utilizing Eqs. (\ref{14}), (\ref{16}) and (\ref{17}), the condition that $v_c$ satisfies can be derived as
\begin{equation}
v_c^{2}-24Q^2+\frac{8^{\omega_q}(3\omega_q+2)(\omega_q+1)\times 9a\omega_q}{v_c^{3\omega_q-1}}=0.\label{18}
\end{equation}
Utilizing  Eqs. (\ref{14}) and (\ref{17}), one can obtain the critical pressure as
\begin{equation}
P_c=\frac{1}{2\pi v_c^2}-\frac{6Q^2}{\pi v_c^4}+\frac{8^{\omega_q}\times 3a\omega_q(2+3\omega_q)}{\pi v_c^{3(1+\omega_q)}}.\label{102}
\end{equation}

 When $\omega_q=-2/3$, Eq. (\ref{18}) can be analytically solved and the corresponding critical quantities can be derived as
\begin{equation}
v_c=2\sqrt{6}Q,\;\;T_c=\frac{\sqrt{6}}{18\pi Q}-\frac{a}{2\pi},\;\;P_c=\frac{1}{96\pi Q^2}.\label{19}
\end{equation}
Comparing Eq. (\ref{19}) with the results of RN-AdS black holes~\cite{Kubiznak}, it is quite interesting to note that the effect of quintessence dark energy is reflected in the critical temperature by adding the second term while the critical specific volume and pressure do not change. When $a=0$, all these critical quantities reduce to those of RN-AdS black holes~\cite{Kubiznak}. When $\omega_q=-2/3$, the third term vanishes in both Eqs. (\ref{18}) and (\ref{102}) just as the case $a=0$. So it is not difficult to explain why the critical volume and pressure do not change.

When $\omega_q\neq-2/3$, we appeal to numerical method for help. To observe the influence of parameters respectively, one can let one parameter vary while keeping others fixed. For specific values of parameters, Eq. (\ref{18}) can be solved numerically by Mathematica programming and the critical specific volume can be derived. Then the critical temperature can be obtained through Eq. (\ref{17}) and the critical pressure can be derived through Eq. (\ref{102}).

Firstly, we set the normalization factor $a=0.1$ and the state parameter $\omega_q=-0.9$. The corresponding critical physical quantities are shown in Table \ref{tb1}. From Table \ref{tb1}, one can find that the critical specific volume $v_c$ increases with $Q$ while the critical Hawking temperature $T_c$ and critical pressure $P_c$ decrease with $Q$. This result is qualitatively in accord with that of the charged AdS black hole without quintessence surrounding it~\cite{Kubiznak}. However, the ratio $\frac{P_cv_c}{T_c}$ does not keep constant as the charged AdS black hole without quintessence surrounding it does~\cite{Kubiznak}. In contrast, the ratio increases with $Q$. This phenomenon reflects again the effects of the dark energy.

\begin{table}[!h]
\tabcolsep 0pt
\caption{Critical physical quantities for $\omega_q=-0.9,a=0.1$}
\vspace*{-12pt}
\begin{center}
\def\temptablewidth{0.5\textwidth}
{\rule{\temptablewidth}{1pt}}
\begin{tabular*}{\temptablewidth}{@{\extracolsep{\fill}}ccccc}
$Q$ & $T_c$ &$v_c$ &$P_c$ &$\frac{P_cv_c}{T_c}$ \\   \hline
   0.5 &0.07925 &       2.40338& 0.02036& 0.618  \\
     1.0  & 0.03149&      4.63276 & 0.00911& 1.340  \\
     1.5  &0.01344        & 6.65598& 0.00665 & 3.294   \\
         2.0  &0.00311&        8.49200& 0.00561 & 15.318
       \end{tabular*}
       {\rule{\temptablewidth}{1pt}}
       \end{center}
       \label{tb1}
       \end{table}

Secondly, we set the normalization factor $a=0.1$ and the charge $Q=1$ and let the state parameter $\omega_q$ vary from $-0.4$ to $-0.9$. The corresponding critical physical quantities are shown in Table \ref{tb2}. The critical specific volume $v_c$ increases with $\omega_q$. Both the critical Hawking temperature $T_c$ and the critical pressure $P_c$ first decrease with $\omega_q$ and then increase with it. The ratio $\frac{P_cv_c}{T_c}$ decreases with $\omega_q$.

\begin{table}[!h]
\tabcolsep 0pt
\caption{Critical physical quantities for $Q=1,a=0.1$}
\vspace*{-12pt}
\begin{center}
\def\temptablewidth{0.5\textwidth}
{\rule{\temptablewidth}{1pt}}
\begin{tabular*}{\temptablewidth}{@{\extracolsep{\fill}}ccccc}
$\omega_q$ & $T_c$ &$v_c$ &$P_c$ &$\frac{P_cv_c}{T_c}$ \\   \hline
   -0.4 &0.03510 &       5.17694& 0.00259& 0.382  \\
     -0.5  & 0.03201&      5.13592 & 0.00256& 0.410  \\
     -0.6  &0.02898        & 5.01322& 0.00283 & 0.490   \\
         -0.7  &0.02688&        4.83892& 0.00369 & 0.664   \\
         -0.8  &0.02707&        4.68218& 0.00558 & 0.966   \\
             -0.9  &0.03149&        4.63276& 0.00911& 1.340
       \end{tabular*}
       {\rule{\temptablewidth}{1pt}}
       \end{center}
       \label{tb2}
       \end{table}

Thirdly, we set the charge $Q=1$ and the state parameter $\omega_q=-0.9$ and investigate the effect of the normalization factor $a$. The corresponding critical physical quantities are shown in Table \ref{tb3}. Both the critical specific volume $v_c$ and the critical Hawking temperature $T_c$ decrease with $a$ while the critical pressure $P_c$ increases with $a$. The ratio $\frac{P_cv_c}{T_c}$ increases with $a$.

\begin{table}[!h]
\tabcolsep 0pt
\caption{Critical physical quantities for $\omega_q=-0.9,Q=1$}
\vspace*{-12pt}
\begin{center}
\def\temptablewidth{0.5\textwidth}
{\rule{\temptablewidth}{1pt}}
\begin{tabular*}{\temptablewidth}{@{\extracolsep{\fill}}ccccc}
$a$ & $T_c$ &$v_c$ &$P_c$ &$\frac{P_cv_c}{T_c}$ \\   \hline
   0.1 &0.03149 &       4.63276& 0.00911& 1.340  \\
     0.2  & 0.02007&      4.43598 & 0.01500& 3.316  \\
     0.3  &0.00895        & 4.28020& 0.02095 & 10.020
       \end{tabular*}
       {\rule{\temptablewidth}{1pt}}
       \end{center}
       \label{tb3}
       \end{table}

\begin{figure*}
\centerline{\subfigure[]{\label{1a}
\includegraphics[width=8cm,height=6cm]{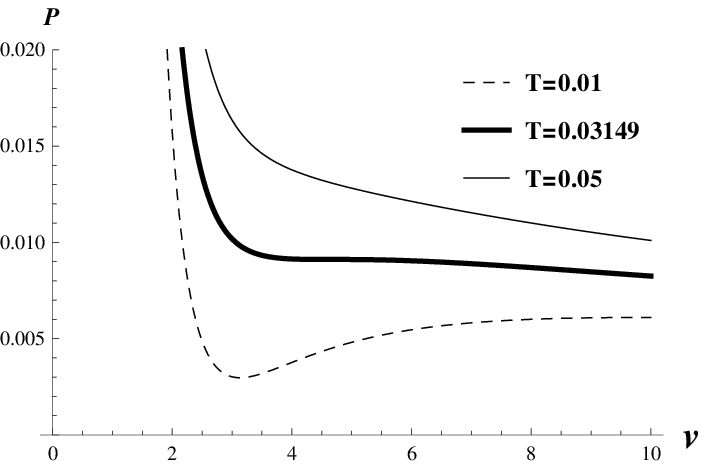}}
\subfigure[]{\label{1b}
\includegraphics[width=8cm,height=6cm]{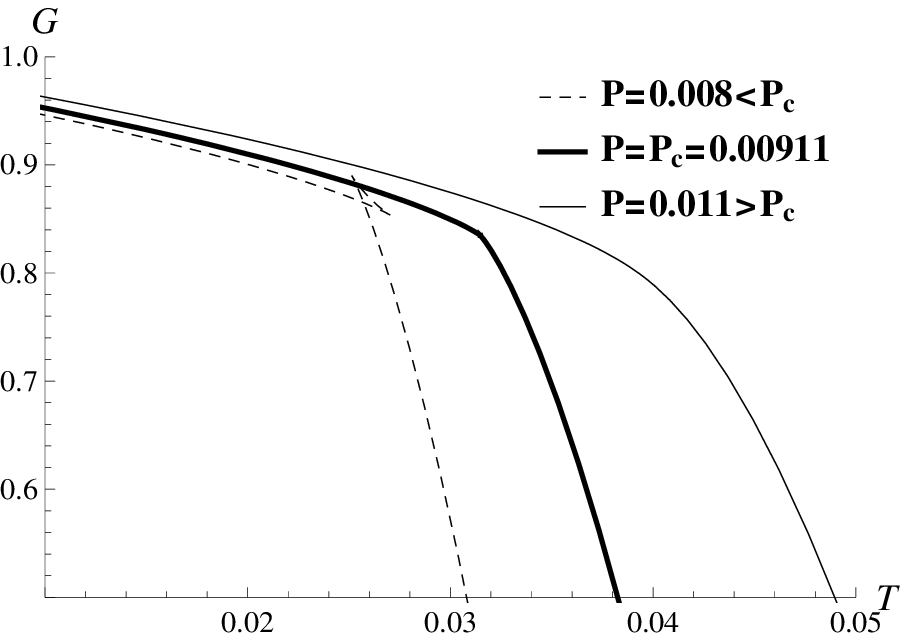}}}
 \caption{(a) $P$ vs. $v$ for $\omega_q=-0.9, Q=1, a=0.1\;\;\;$(b) $G$ vs. $T$ for $\omega_q=-0.9, Q=1, a=0.1$} \label{fg1}
\end{figure*}

To observe the behavior of $P$ more intuitively, its curve is plotted under different temperature. As shown in Fig.\ref{1a}, the isotherm corresponding to the temperature less than the critical temperature can be divided into three branches. Both the small radius branch and the large
radius branch are stable while the medium radius branch is unstable. There is phase transition between
the small black hole and the large black hole. To understand this phase transition more deeply, one can analyze the behavior of Gibbs free energy.

In the extended phase space, the mass is interpreted as enthalpy. So the Gibbs free energy can be derived as
\begin{equation}
G=H-TS=M-TS=\frac{3Q^2}{4r_+}+\frac{r_+}{4}-\frac{2P\pi r_+^3}{3}-\frac{a(2+3\omega_q)}{4r_+^{3\omega_q}}.\label{20}
\end{equation}
The behavior of Gibbs free energy is depicted in Fig.\ref{1b}. The classical swallow tail phenomenon observed below the critical temperature implies that the existence of first order phase transition.
%%%%%%%%%%%%%%%%%%%%%%%%%%%%%%%%%%%%%%%%%%%%%%%%%%%%%%%%%%%%%%%%%%%%%%%%%%%%%
%%%%%%%%%%%%%%%%%%%%%%%%%%%%%%%%%%%%%%%%%%%%%%%%%%%%%%%%%%%%%%%%%%%%%%%%%%%%%%%%

\section{Critical exponents of the RN-AdS black hole surrounded by quintessence}
\label{Sec4}
Critical exponents are often used to describe the critical behavior near the critical point. It is convenient to introduce the following notations
\begin{equation}
t=\frac{T}{T_c}-1,\;\;\epsilon=\frac{v}{v_c}-1,\;\;p=\frac{P}{P_c}.\label{21}
\end{equation}
The definitions of critical exponents are as follows.
\begin{eqnarray}
C_v&\propto&|t|^{-\alpha},\label{22}\\
\eta&\propto&|t|^{\beta},\label{23}\\
\kappa_T&\propto&|t|^{-\gamma},\label{24}\\
|P-P_c|&\propto&|v-v_c|^{\delta}.\label{25}
\end{eqnarray}

$\alpha$ describes the behavior of specific heat when the volume is fixed. From Eq. (\ref{6}), it is not difficult to draw the conclusion that the entropy $S$ is independent of the Hawking temperature $T$. So
\begin{equation}
C_V=T\left(\frac{\partial S}{\partial T}\right)_V=0,\label{26}
\end{equation}
from which one can derive that $\alpha=0$.

  $\beta$ characterizes the behavior of the order parameter $\eta$. Near the critical point, the equation of state can be expanded into
  \begin{equation}
p=1+b_{10}t+b_{01}\epsilon+b_{11}t\epsilon+b_{02}\epsilon^2+b_{03}\epsilon^3+O(t\epsilon^2,\epsilon^4),\label{27}
\end{equation}
where
\begin{eqnarray}
b_{01}&=&b_{02}=0,\label{28}\\
b_{10}&=&\frac{T_c}{P_cV_c},\label{29}\\
b_{11}&=&-\frac{T_c}{P_cV_c},\label{30}\\
b_{03}&=&\frac{24T_c}{P_cv_c}-\frac{18}{\pi P_cv_c^2}+\frac{8^{\omega_q}\times9a\omega_q(-10+17\omega_q+36\omega_q^2+9\omega_q^3)}{\pi P_cv_c^{3(1+3\omega_q)}}.\label{31}
\end{eqnarray}
Since the pressure keeps constant during the phase transition, one can obtain
 \begin{equation}
1+b_{10}t+b_{11}t\epsilon_l+b_{03}\epsilon_l^3=1+b_{10}t+b_{11}t\epsilon_s+b_{03}\epsilon_s^3,\label{32}
\end{equation}
where the subscripts $l$ and $s$ denote the large black hole and small black hole respectively.
On the other hand, one can apply Maxwell's equal area law
 \begin{equation}
\int^{\epsilon_s}_{\epsilon_l}\epsilon \frac{dp}{d\epsilon}d \epsilon=0,\label{33}
\end{equation}
where
\begin{equation}
\frac{dp}{d\epsilon}=b_{11}t+3b_{03}\epsilon^2,\label{34}
\end{equation}
and obtain
 \begin{equation}
 b_{11}t(\epsilon^2_s-\epsilon^2_l)+\frac{3}{2} b_{03}(\epsilon^4_s-\epsilon^4_l)=0.\label{35}
\end{equation}
From Eqs. (\ref{32}) and (\ref{35}), one can obtain
\begin{equation}
\epsilon_l=-\epsilon_s=\sqrt{\frac{-b_{11}t}{b_{03}}}.\label{36}
\end{equation}
So
\begin{equation}
\eta=v_l-v_s=v_c(\epsilon_l-\epsilon_s)=2v_c\epsilon_l\propto\sqrt{-t},\label{37}
\end{equation}
yielding $\beta=1/2$.

$\gamma$ describes the behavior of isothermal compressibility $\kappa_T$, which can be derived as
\begin{equation}
\kappa_T=\left.-\frac{1}{v}\frac{\partial v}{\partial P}\right|_{v_c}\propto \left.-\frac{1}{\frac{\partial p}{\partial \epsilon}}\right|_{\epsilon=0}=-\frac{1}{b_{11}t},\label{38}
\end{equation}
which yields $\gamma=1$.

$\delta$ characterizes the behavior of the critical isotherm corresponding to $T=T_c$. Substituting $t=0$ into Eq. (\ref{31}), one can obtain
\begin{equation}
p-1=b_{03}\epsilon^3,\label{39}
\end{equation}
yielding $\delta=3$.

From the above derivations, we can see clearly that four critical exponents are exactly the same as those obtained before for charged AdS black holes. This implies that quintessence dark energy does not change the critical exponents.
\section{Conclusions}
\label{Sec5}
    By treating the cosmological constant as thermodynamic pressure, we study the thermodynamics of the RN-AdS black hole surrounded by quintessence in the extended phase space. It is quite interesting to note that the existence of quintessence dark energy does not affect the expressions of both the entropy and the thermodynamic volume of RN-AdS black holes. Quintessence dark energy does not affect the existence of small/large black hole phase transition either. It is shown that the isotherm corresponding to the temperature less than the critical temperature can be divided into three branches. Both the small radius branch and the large radius branch are stable while the medium radius branch is unstable. And the classical swallow tail phenomenon implying the existence of first order phase transition is also observed below the critical temperature in the Gibbs free energy graph.

    However, quintessence dark energy affects the critical physical quantities near the critical point. For the case $\omega_q=-2/3$, it is possible to derive analytic expressions of critical physical quantities. It is quite interesting to note that the critical specific volume and pressure are exactly
the same as those of RN-AdS black holes~\cite{Kubiznak} while the critical temperature gain an extra term due to the existence of quintessence dark energy. For cases $\omega_q\neq-2/3$, we appeal to numerical method for help. Firstly, we fix the normalization factor $a$ and the state parameter $\omega_q$. The critical specific volume $v_c$ increases with $Q$ while the critical Hawking temperature $T_c$ and critical pressure $P_c$ decrease with $Q$. This result is qualitatively in accord with that of the charged AdS black hole without quintessence surrounding it~\cite{Kubiznak}. However, the ratio $\frac{P_cv_c}{T_c}$ does not keep constant as the charged AdS black hole without quintessence surrounding it~\cite{Kubiznak} does. In contrast, the ratio increases with $Q$. This phenomenon reflects again the effects of the dark energy. Secondly, we observe the impact of the state parameter $\omega_q$. The critical specific volume $v_c$ increases with $\omega_q$. Both the critical Hawking temperature $T_c$ and the critical pressure $P_c$ first decrease with $\omega_q$ and then increase with it. The ratio $\frac{P_cv_c}{T_c}$ decreases with $\omega_q$. Thirdly, we focus on the impact of the normalization factor $a$. Both the critical specific volume $v_c$ and the critical Hawking temperature $T_c$ decrease with $a$ while the critical pressure $P_c$ increases with $a$. The ratio $\frac{P_cv_c}{T_c}$ increases with $a$.

In the end, we would like to discuss more on the findings of this Letter.  On the one hand, the critical exponents calculated in this Letter are the same as those of other black holes in former literatures, suggesting again that the critical behaviour of black holes in the extended space also obeys the mean field theory and that AdS black holes may belong to the same universality class as Van der Vaals liquid-gas systems. On the other hand, the ratio $\frac{P_cv_c}{T_c}$ does not keep constant as the RN-AdS black hole does. In fact, it depends on the state parameter $\omega_q$ and the normalization factor $a$. This reflects the influence of the quintessence dark energy. This result is quite natural for it has been shown in former literatures that the ratio $\frac{P_cv_c}{T_c}$ relies on some characteristic quantities of black hole spacetimes, such as dimensionality~\cite{Belhaj99}, Born-Infeld parameter~\cite{Gunasekaran}, Lovelock parameter~\cite{Wenbiao3} etc. The results in this Letter show the impact of dark energy, not only on the first law but also on the critical physical quantities. Hence it would be of great importance to investigate the implications of black
holes surrounded by quintessence as a dark energy model. And it should be further probed in the future.

 \section*{Acknowledgements}
The author would like to express sincere gratitude to the anonymous referee for his enlightening suggestions which have help improve the quality of this Letter greatly. The author also appreciates the editor's time and patience on this Letter.


\begin{thebibliography}{99}


\bibitem{Chamblin1}
 A. Chamblin, R. Emparan, C.V. Johnson and R.C. Myers, Charged AdS black Holes and catastrophic holography, Phys. Rev. D 60, 064018(1999).

\bibitem{Chamblin2}
A. Chamblin, R. Emparan, C.V. Johnson and R.C. Myers, Holography,
thermodynamics and fluctuations of charged AdS black holes, Phys.
Rev.D 60, 104026(1999).

\bibitem{Kastor}
 D. Kastor, S. Ray, and J. Traschen, Enthalpy and the mechanics of AdS black holes, Class. Quant. Grav. 26, 195011(2009).

\bibitem{Dolan1}
  B. Dolan, The cosmological constant and the black hole equation of state, Class. Quant. Grav. 28, 125020(2011).

\bibitem{Dolan2}
B. P. Dolan, Pressure and volume in the first law of black hole thermodynamics, Class. Quant. Grav. 28, 235017(2011).

\bibitem{Dolan3}
 B. P. Dolan, Compressibility of rotating black holes, Phys. Rev. D 84, 127503(2011).

 \bibitem{Cvetic}
M. Cvetic, G. Gibbons, D. Kubiznak, and C. Pope, Black hole enthalpy and an entropy inequality for the thermodynamic volume, Phys. Rev. D 84, 024037(2011).

\bibitem{Kubiznak}
D. Kubiz\v{n}\'{a}k and  R. B. Mann, $P$-$V$ criticality of charged AdS black holes, JHEP 1207, 033(2012).

\bibitem{Gunasekaran}
S. Gunasekaran, R. B. Mann and D. Kubiz\v{n}\'{a}k, Extended phase space thermodynamics for charged and rotating black holes and Born-Infeld vacuum polarization, JHEP 1211, 110(2012).

\bibitem{Belhaj99}
A. Belhaj, M. Chabab, H. El Moumni and M. B. Sedra, On thermodynamics of AdS black holes in arbitrary dimensions, Chin. Phys. Lett. 29, 100401(2012) .

\bibitem{Chen}
S. Chen, X. Liu, C. Liu and J. Jing, $P$-$V$ criticality of AdS black hole in $f(R)$ gravity, Chin. Phys. Lett. 30, 060401(2013).

\bibitem{Hendi}
 S. H. Hendi and M. H. Vahidinia, Extended phase space thermodynamics and $P$-$V$ criticality of black holes with nonlinear source, Phys. Rev. D 88, 084045(2013).

\bibitem{Spallucci}
E. Spallucci and A. Smailagic, Maxwell's equal area law for charged Anti-deSitter black holes, Phys. Lett. B 723, 436-441(2013).

\bibitem{Zhao}
R. Zhao, H. H. Zhao, M. S. Ma and L. C. Zhang, On the critical phenomena and thermodynamics of charged topological dilaton AdS black holes, Eur. Phys. J. C 73, 2645(2013).

\bibitem{Altamirano1}
 N. Altamirano, D. Kubiz\v{n}\'{a}k and R. Mann, Reentrant phase transitions in rotating AdS black holes, Phys. Rev. D 88, 101502(2013).

\bibitem{Cai98}
R. G. Cai, L. M. Cao, L. Li and  R. Q. Yang, $P$-$V$ criticality in the extended phase space of Gauss-Bonnet black holes in AdS space, JHEP 1309, 005(2013).


 \bibitem{Wenbiao1}
J. X. Mo and W. B. Liu, Ehrenfest scheme for $P$-$V$ criticality in the extended phase space of black holes, Phys. Lett. B 727, 336-339 (2013).

\bibitem{Wenbiao2}
J. X. Mo, X. X. Zeng, G. Q. Li, X. Jiang, W. B. Liu, A unified phase transition picture of the charged topological black hole in Ho\v{r}ava-Lifshitz gravity, JHEP 1310, 056(2013).

\bibitem{Altamirano2}
 N. Altamirano, D. Kubiz\v{n}\'{a}k, R. Mann and Z. Sherkatghanad, Kerr-AdS analogue of tricritical point and solid/liquid/gas phase transition, Class. Quant. Grav. 31(4), 042001(2014).


\bibitem{Decheng}
D. C. Zou, S. J. Zhang and B. Wang, Critical behavior of Born-Infeld AdS black holes in the extended phase space thermodynamics, Phys. Rev. D 89, 044002 (2014).

 \bibitem{Wenbiao3}
J. X. Mo and W. B. Liu, $P$-$V$ Criticality of topological black holes in Lovelock-Born-Infeld gravity, Eur. Phys. J. C 74, 2836(2014).

 \bibitem{Wenbiao4}
J. X. Mo and W. B. Liu, Ehrenfest scheme for $P$-$V$ criticality of higher dimensional charged black holes, rotating black holes and Gauss-Bonnet AdS black holes, Phys. Rev. D 89, 084057(2014).


\bibitem{Altamirano3}
 N. Altamirano, D. Kubiz\v{n}\'{a}k, R. Mann and Z. Sherkatghanad, Thermodynamics of rotating black holes and black rings: phase transitions and thermodynamic volume, Galaxies 2, 89-159(2014).

\bibitem{Mengsen1}
M. S. Ma, H. H. Zhao, L. C. Zhang and R. Zhao, Existence condition and phase transition of Reissner-Nordstr\"{o}m-de Sitter black hole, Int. J. Mod. Phys. A 29, 1450050(2014).



\bibitem{shaowenwei}
S. W. Wei and Y. X. Liu, Triple points and phase diagrams in the extended phase space of charged Gauss-Bonnet black holes in AdS space, arXiv:1402.2837.

\bibitem{Kubiznak2}
D. Kubiz\v{n}\'{a}k and  R. B. Mann, Black hole chemistry, arXiv:1404.2126.

\bibitem{Mengsen2}
L. C. Zhang, M. S. Ma, H. H. Zhao and R. Zhao, Thermodynamics of phase transition in higher dimensional Reissner-Nordstr\"{o}m-de Sitter black hole, arXiv:1403.2151.


\bibitem{Dolan4}
B. P. Dolan, Thermodynamic stability of asymptotically anti-de Sitter rotating black holes in higher dimensions, arXiv:1403.1507.

\bibitem{Decheng2}
D. C. Zou, Y. Liu and B. Wang, Critical behavior of charged Gauss-Bonnet AdS black holes in the grand canonical ensemble, arXiv:1404.5194.



\bibitem{Bachall}
N. A. Bachall, J. P. Ostriker, S. Perlmutter and P. J. Steinhardt, The cosmic triangle: Revealing the state of the universe, Science 284, 1481(1999).

\bibitem{Perlmutter}
S. J. Perlmutter et al, Measurements of Omega and Lambda from 42 High-Redshift Supernovae, Astrophys. J. 517, 565(1999).

\bibitem{Sahni}
V. Sahni and A. A. Starobinsky, The case for a positive cosmological Lambda-term, Int. J. Mod. Phys. D 9, 373(2000).


\bibitem{Fujii}
Y. Fujii, Origin of the gravitational constant and particle masses in a scale-invariant scalar-tensor theory, Phys. Rev. D 26, 2580(1982).

\bibitem{Ford}
L. H. Ford, Cosmological-constant damping by unstable scalar fields, Phys. Rev. D 35, 2339(1987).

\bibitem{Wetterich}
C. Wetterich, Cosmology and the fate of dilatation symmetry, Nucl. Phys B. 302, 668(1988).


\bibitem{Ratra}
B. Ratra and P. J. E. Peebles, Cosmological consequences of a rolling homogeneous scalar field, Phys. Rev. D 37, 3406 (1988).


\bibitem{Tsujikawa}
Shinji Tsujikawa, Quintessence: A Review, Class. Quant. Grav. 30, 214003(2013).

\bibitem{Kiselev}
V. V. Kiselev, Quintessence and black holes, Class. Quant. Grav. 20, 1187-1198(2003).

\bibitem{songbai1}
S. B. Chen and J. L. Jing, Quasinormal modes of a black hole surrounded by quintessence,  Class. Quant. Grav. 22, 4651-4657(2005).

\bibitem{guiyuanxing}
Y. Zhang, Y. X. Gui and F. L. Li, Quasinormal modes of a Schwarzschild black hole surrounded by quintessence: Electromagnetic perturbations, Gen. Rel. Grav. 39, 1003-1010(2007).

\bibitem{guiyuanxing2}
Y. Zhang and Y. X. Gui, Quasinormal modes of a Schwarzschild black hole surrounded by quintessence, Class. Quant. Grav. 23, 6141-6147(2006) .

\bibitem{songbai2}
S. Chen, B. Wang and R. Su , Hawking radiation in a d-dimensional static spherically-symmetric black Hole surrounded by quintessence, Phys. Rev. D 77, 124011(2008).

\bibitem{Kuriakose}
N. Varghese and V. C. Kuriakose, Massive Charged Scalar Quasinormal Modes of Reissner-Nordstrom Black Hole Surrounded by Quintessence, Gen. Rel. Grav. 41, 1249-1257(2009).

\bibitem{Saleh3}
M. Saleh, B. T. Bouetou and T. C. Kofane, Quasi-normal modes of gravitational perturbation around a Reissner-Nordstroem black hole surrounded by quintessence, Chin. Phys. Lett. 26, 109802(2009).

\bibitem{guiyuanxing3}
C. Y. Wang, Y. Zhang, Y. X. Gui and J. B. Lu, Dirac quasinormal modes of Reissner-Nordstroem black hole surrounded by quintessence, Commun. Theor. Phys. 53, 882-888(2010).

\bibitem{guqiang}
G. Q. Li and S. F. Xiao, State equations for massless spin fields in static spherical spacetime filled with quintessence, Gen. Rel. Grav. 42, 1719-1726(2010).

\bibitem{Saleh2}
M. Saleh, B. T. Bouetou and T. C. Kofane, Quasinormal modes and Hawking radiation of a Reissner-Nordstroem black hole surrounded by quintessence, Astrophys. Space Sci. 333, 449-455(2011).

\bibitem{yihuan}
Y. H. Wei and Z. H. Chu, Thermodynamic properties of a Reissner-Nordstroem quintessence black hole, Chin. Phys. Lett. 28, 100403(2011).

\bibitem{Fernando}
S. Fernando, Schwarzschild black hole surrounded by quintessence: Null geodesics, Gen. Rel. Grav. 44, 1857-1879(2012).

\bibitem{Saleh}
B. B. Thomas and M. Saleh, Thermodynamics and phase transition of the Reissner-Nordstroem black hole surrounded by quintessence, Gen. Rel. Grav. 44, 2181-2189(2012).

\bibitem{sonbai3}
S. Chen, Q. Pan and J. Jing, Holographic superconductors in quintessence AdS black hole spacetime, Class. Quant. Grav. 30, 145001(2013).

\bibitem{Rodrigues}
M. A. A\"{\i}nou and M. E. Rodrigues, Thermodynamical, geometrical and Poincar¨¦ methods for charged black holes in presence of quintessence, JHEP 1309, 146(2013) .

\bibitem{Kuriakose2}
R. Tharanath, N. Varghese and V.C. Kuriakose, Thermodynamics and Spectroscopy of Schwarzschild black hole surrounded by Quintessence, Mod. Phys. Lett. A 28, 1350003(2013).

\bibitem{Kuriakose3}
R. Tharanath and V.C. Kuriakose, Phase transition, Quasinormal modes and Hawking radiation of Schwarzschild black hole in Quintessence field, Mod. Phys. Lett. A 29, 1450057(2014).

\end{thebibliography}
\end{document}